# Toward Transdisciplinary Approaches to Audio Deepfake Discernment

Vandana P. Janeja[1,2], Christine Mallinson[3]

**Abstract:** This perspective calls for scholars across disciplines to address the challenge of audio deepfake detection and discernment through an interdisciplinary lens across Artificial Intelligence methods and linguistics. With an avalanche of tools for the generation of realistic-sounding fake speech on one side, the detection of deepfakes is lagging on the other. Particularly hindering audio deepfake detection is the fact that current AI models lack a full understanding of the inherent variability of language and the complexities and uniqueness of human speech. We see the promising potential in recent transdisciplinary work that incorporates linguistic knowledge into AI approaches to provide pathways for expert-in-the-loop and to move beyond expert agnostic AI-based methods for more robust and comprehensive deepfake detection.

**Introduction**

Fifteen years ago, *The Fourth Paradigm (1)* emphasized the significance of the data revolution that has now thoroughly permeated our global society. While the benefits and opportunities have been tremendous, the data revolution and Artificial Intelligence (AI) advances, in particular generative AI led advances, have also led to serious societal threats and risks *(2)*. Alarms have particularly been raised over an omnipresent flood of false content, including AI-generated fake content (i.e. deepfakes) that threaten our ability to separate fact from fiction *(23)*. While all efforts to combat deepfakes are critically important–from improving detection systems to passing widespread legislation, to attention to ethical and responsible AI–we assert that deepfake detection techniques must also move away from relying on expert agnostic AI-based detection algorithms to prevent harm. Through human-in-the-loop methods that incorporate complexities of the human experience–for example, the structure and features of language in use–we can improve discernment and detection to more comprehensively address the deepfake challenge.

**An Arms Race for Authenticity**

In 2023, Merriam Webster declared "authenticity" to be the word of the year *(3)*–an indicator that, in the current AI era, the ability to differentiate between what is real and unreal is increasingly difficult. False images masquerade as celebrities, robocalls impersonate political candidates, and readily available apps can manipulate anyone's voice in seconds in authorized *(4)* or unauthorized *(5)* ways. The trustworthiness of how humans communicate has been especially undermined and disrupted by deepfakes–images, videos, texts, and audio content that has been synthetically generated or manipulated using AI methods. These hyper-realistic manipulations have been used as powerful vehicles for deception, fraud, threat, misinformation and disinformation, and social turmoil *(6, 7)*.

---

[1] Department of Information Systems, University of Maryland, Baltimore County (UMBC); (Baltimore, Maryland, U.S.A.).
[2] Corresponding author. Email: vjaneja@umbc.edu; Both authors contributed equally.
[3] Language, Literacy, and Culture Program, University of Maryland, Baltimore County (UMBC); (Baltimore, Maryland, U.S.A.).

Two serious gaps are hindering researchers' ability to comprehensively combat the rapid rise in deepfake technology. First, there is a lack of sustained attention to audio. Audio cues are a foundational building block across deepfake technology, including in video deepfakes. Audio is also often used for authentication, such as with smart devices and smart home systems, where spoofed audio can lead to unauthorized access, causing security risks. To date, however, audio deepfakes have been slower to capture the attention of researchers. As shown in a review of over 140 scientific papers on deepfakes *(8)*, research has primarily focused on text and video deepfakes, some on image deepfakes, and far less on audio. Second, current efforts to curtail deepfakes have focused primarily on labeling content as fake and then refining AI algorithms to "catch" it. This overreliance on generative AI as being the main solution to, but also the main source of, deepfakes leads to a vicious cycle: spoofed content is generated, the research community puts out a new AI algorithm to detect it, and the adversary generates better deepfakes. AI algorithms can only iterate with new training datasets, while adversaries improve with small unique samples of data. In short, "the battle between deepfake creation and detection is an arms race" *(6)*.

**Generative AI: Boon and Bane**
The present impasse can be traced in the history of generative versus discriminative AI from the early 2000s *(9)*. At that point, the machine learning world was leading with discriminative algorithms, such as support vector machines. Proponents of this technology were happily claiming to have solved classification problems, as with businesses who wanted to predict whether an unknown person would buy goods or not, based on features from labeled customers who had bought such goods in the past. But a lack of labeled data–and, in particular, *good* labeled data–led to the (re)emergence of generative AI methods. By sampling distributions of known samples, new samples could be generated. The applications are endless. We now have the potential to use generative AI models to generate anything with a distribution that can be modeled–which is virtually anything, given some known samples.

The data lifecycle is precarious, however. In its traditional conceptualization, there has been an emphasis on the need to embed ethical thinking throughout *(10)*, so that as data moved from collection to integration, selection, and preprocessing, it would be carefully vetted by humans to limit bias in the models. Moreover, in a traditional view of the data lifecycle, patterns detected from data must be evaluated for ground truths, model verifications, and business cases where they are implemented. For example, in 2007, Jim Gray expounded, "after you have captured the data, you need to curate it before you can start doing any kind of data analysis" *(1)*.

Unfortunately, this ideal model of the data lifecycle has not been robustly implemented and, over time, has resulted in a leaky process. Known bias has crept into models based on faulty input data, and few checks and balances are available to prevent the continued suffusion of bias into the massive amounts of generated data that currently exist. At present, ever-more data is being haphazardly thrown into models, with known biases becoming embedded into applications in ways that are difficult, if not near-impossible, to disrupt *(11)*.

Issues surrounding validity are further aggravated by methods that rely upon replacing data from the real world with information derived from generative models–not only generated data but also generated code, all of which carry the risk of unintended consequences. Although models may be following a mathematical or probabilistic method of generative data from sample distributions, they cannot account for the inherent complexities in situations where data cannot be modeled based only on a few samples. Such models also suffer from lack of verifiability and interpretability. Put simply, we cannot accurately model the complexity of the dynamics of the ocean based on a few sampled waves. These insights inform audio deepfake detection as a use case, since most detection algorithms do not perform well because they are unable to fully capture the uniqueness, variability, and complexity of human language.

**The Importance of Understanding Language in AI-Based Language Models**

AI-based language models (such as Natural Language Processing (NLP), Large Language Models (LLM)), are facing a crisis of validity along multiple dimensions. First, existing models often incorrectly generalize, based on reductive training data derived primarily from one language, English. As computational linguist Emily Bender notes, "*natural language* is not a synonym for *English*" *(12)*. Most training data and major research in NLP focuses so overwhelmingly on English to the point that it is viewed as the default. Yet, English should not be viewed, or used, as a stand-in for the world's approximately 7,000 other languages. Among other particularities, English cannot generalize to non-written languages, to signed languages, to languages that use logographic systems such as Chinese or Japanese kanji, or to languages that use non-Latin scripts such as Arabic, Cyrillic, and Hebrew; there is also a vast amount of training data available for English but not for other languages, which skews methods and results *(12)*. As with many other social dimensions, a lack of linguistic diversity in inputs leads to a homogeneity of outputs–including one-dimensional AI detection models with flawed underlying distributions.

Existing AI-based language models also do not capture the inherent variability of language *(13)*. As robustly established by research in linguistics, especially the subfield of sociolinguistics, human language is dynamic, complex, and evolving *(26)*. Language affords us with near-infinite possibilities for using language in creative and innovative ways that are essential for communication, interaction, and meaning making. There are multiple varieties (dialects) of any given language, and within those varieties, researchers can statistically document patterns of usage that align with myriad and intersecting social dimensions and that change and shift over time. Language variation encapsulates the different accents, vocabulary items, grammatical structures, stylistic elements, and other linguistic characteristics that sometimes go unnoticed but, more often than not, become emblematic of people's identity and culture. Though large language models and training data may treat it as such, the reality is that human language is much more than a finite set of features, and it is far from one-size fits all. Understanding the uniqueness and variability of language in use can be an important tool for enhancing deepfake detection models.

**Audio Deepfakes: A Use Case of the Problem and a Solution**

In September 2021, the *New York Times* reported on a story in which the company Ozy Media tried to drive a fraudulent $40 million investment from Goldman Sachs. In a near-final phone call, one of the voices "began to sound strange to the Goldman Sachs team, as though it might have been digitally altered" *(14)*. The situation turned out to be an attempt to impersonate a key executive to close the deal—thwarted only by a listener's keen ear. This incident shines light on a powerful tool at our disposal in confronting audio deepfakes: the ability to identify subtle but still-noticeable cues that a voice sounds "strange."

Bringing human understanding to the audio deepfake challenge is an opportunity that we cannot afford to overlook. There are two main pathways for human-in-the-loop and expert-in-the-loop, both of which avoid the current tendency toward overreliance on domain agnostic AI-based deepfake detection methods. First, AI algorithms need to continue to improve, not by iteratively incorporating more of the same types of training data, but by incorporating human knowledge in the form of domain expertise-informed features. Second, humans ourselves need to improve in our ability to discern fake content, including fake audio, by listening for the unique cues of language. Both pathways would benefit from incorporating a more comprehensive understanding of human language in use into existing and new approaches for deepfake detection and discernment.

*Domain Expertise*

To return to the problem of the current leaky data lifecycle and the need for checks and balances, current models for audio deepfake detection are weak in several key areas: they are highly biased toward English and toward standardized forms of English; they have not been strengthened to attend to linguistic variability (e.g., regional variation, sociocultural variation) *(15)*; and they cannot reliably account for conversational or multi-speaker settings. Indeed, most detection methods fail to achieve high accuracy in general *(8)*. In our observation, some generative models used to generate deepfakes, such as text to speech, only incorporate a handful of acoustic features. Caught in the instant gratification cycle of catching the next spoof, detection models fail to fully take into account the patterned complexities of human language–an expert domain of linguists, who are trained to understand its structure, use, and development. As a result, the underlying distribution of such AI-based deepfake detection models are flawed. Meanwhile, accents, conversational features, and other linguistic details are beginning to be added into synthetic speech *(22)*, increasing the gap between generation and detection, as most models for detecting fake audio fail to incorporate these or other linguistic elements. At the same time, efforts to add language variation into synthetic speech must be approached with a great deal of caution, so as to avoid the harmful potential for linguistic stereotyping and bias.

When the emphasis in machine learning is on getting the computer to 'learn' directly from data, domain expertise becomes devalued *(16)*. On the one hand, experts can bring in their own biases, learned information, and lived experiences; on the other hand, removing the input of experts–who should not only be annotators but also analyzers–directly contributes to the limitations of current fake audio detection systems. Expert knowledge (re)integration is a key part of the solution. As originally set forth, the data lifecycle must ensure that humans are involved throughout its entirety–as experts, as data curators and analyzers, and as those who provide checks and balances to ensure that integrity is not lost to scalability. An example of an expert informed data lifecycle is represented in Figure 1, where data from generative applications is linguistics informed and incorporates domain expertise at all stages including when a deepfake detection is used in practice.

Domain expert knowledge has informed our recent innovative approaches that aim to apply an understanding of the linguistic structure and features of fake versus real audio for audio deepfake detection. In our research *(17, 18)*, we found that augmenting spoken English audio data with expert-informed linguistic annotations of commonly occurring, variable, and distinguishing phonetic and phonological features increased the accuracy of spoofed audio detection significantly in training and testing datasets across the evaluated single and ensemble models. This work indicates a promising new avenue for improving traditional detection approaches. The as-yet relatively unexplored frontier of fake audio detection in multi-speaker conversational settings, such as call center recordings featuring interactions with counterfeit representatives and customers, would likely also benefit from (socio)linguistic insight (*27*).

In another linguistically-informed approach to detecting fake speech, researchers drew upon articulatory phonetics and fluid dynamics to reconstruct the shape of a vocal tract based on the acoustic properties of an audio signal–in other words, identifying fake audio by determining whether it was possible for a given speech sample to have been produced by a human vocal tract *(19)*. While accurate, this approach requires technical knowledge of fluid dynamics in addition to deep learning and is not a usable solution for the everyday listener who may encounter fake audio "in the wild"—that is, on social media, over the phone, and so on. There is thus a clear need to further develop non AI-based strategies for deepfake detection—including to invest in discernment by humans.

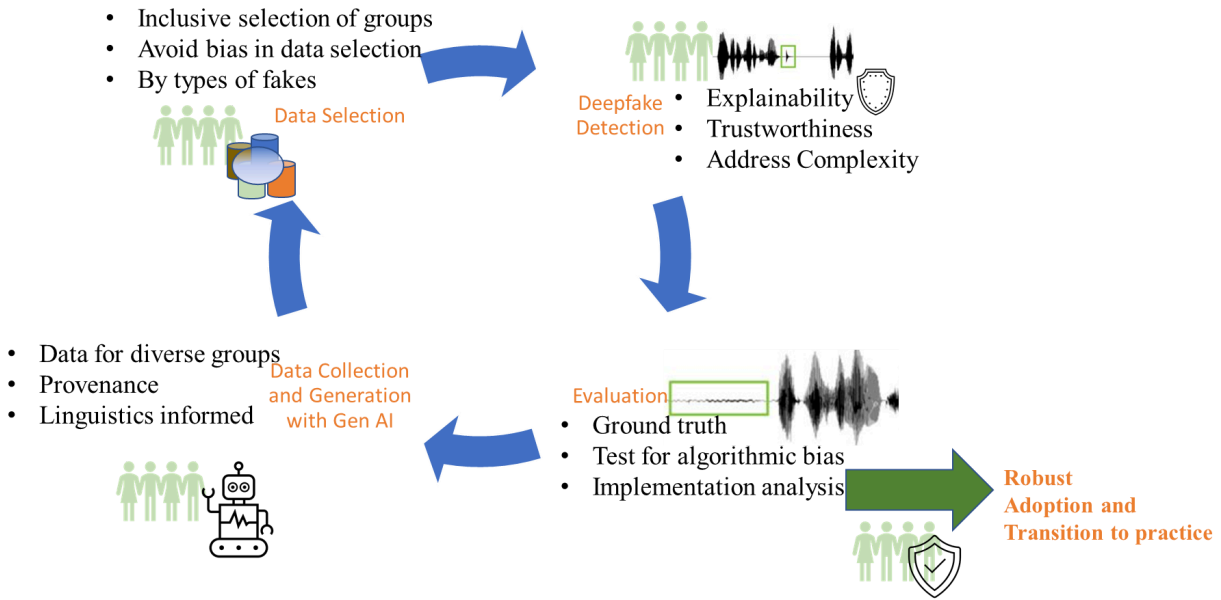

*Fig. 1: Generative AI data lifecycle for deep fake detection - embedding the expert input at various points.*

*Human Discernment*

The ability for humans to react to potentially fake audio in real time is a pressing need, given the speed with which new and sophisticated methods of deepfake generation are being developed and can evade algorithmic detection, and also particularly given the constant bombardment of content. Approaches that rely on individual acoustic features can lose extraneous–but often valuable–information from an audio clip and tend to perform poorly on unseen real-world data *([24](#))*. By contrast, human listeners are able to listen holistically, as did the listener from Goldman Sachs in the case described above, who presumably was able to take into account both linguistic information and non-linguistic elements of the sound signal to question the audio's veracity.

Thus far, few studies have sought to measure human detection capabilities *([7](#))*. Such efforts would improve efforts to develop *inoculation strategies* that users can learn and employ to help protect them from being deceived by manipulated content *([20](#))*. Thus far, however, studies generally have found that humans lack strong perceptual awareness of audio deepfakes and are fairly unreliable at discerning them, even after completing informational training *([21](#))*. Typically, programs designed to train individuals to spot deepfakes either focus on increasing their awareness of deepfakes or coaching them to spot visual cues, such as irregularities in images or videos, leaving untapped the potential of listening for auditory cues. In our pilot studies, we have found that training students to listen for specific characteristics of spoken English as potential "tells" of real versus fake speech increased their certainty in spotting deepfakes *([25](#))*. Guiding humans to listen for audio *and* visual cues would allow individuals to more holistically use all available information to help judge whether media content is authentic.

More research grounded in behavioral and education sciences could also help develop better methods for designing effective deepfake discernment training.

**Conclusion**

Existing and innovative interventions point to the need for combined approaches that capitalize on the potential for improving discernment and detection approaches to combat deepfakes. Such work necessitates collaboration among scholars across disciplines, such as linguists, whose expertise is highly relevant yet who have not fully been included in high-level efforts to detect deepfakes so far. The more we can train humans and scientific models to incorporate the complexities of human language, the more accurate audio deepfake detection will be. The more we can explore and invest in interdisciplinary and transdisciplinary approaches overall, the better positioned we will be to use all tools at our disposal to avert deepfake doomsday scenarios. If teams of scholars and practitioners across fields work together to keep humans in the loop as we model the complexity of our social world, there is great potential to take innovative leaps forward in generative AI research.

**Acknowledgments:** We gratefully acknowledge the contributions of our student team members, in particular Zahra (Sara) Khanjani and Chloe Evered, which have shaped the thinking in our paper. Authors would like to acknowledge funding from NSF award #2210011 and # 2346473.